\documentclass[journal]{IEEEtran}

\usepackage{hyperref}

\usepackage{soul}
\usepackage{array}
\usepackage{multicol}
\usepackage{multirow}
\usepackage{xfrac}
\usepackage{graphicx}
\usepackage{amsmath}
\usepackage{amssymb}
\usepackage{capt-of}

\usepackage{adjustbox}
\usepackage{tikz}
\usepackage{xcolor}
\usepackage[switch]{lineno}
\definecolor{FAUblue1}{RGB}{0,56,101}

\hypersetup{
	colorlinks,
	linkcolor={FAUblue1},
	citecolor={FAUblue1},
	urlcolor={FAUblue1}
}

\DeclareMathOperator*{\argmin}{argmin}
\renewcommand\vec{\mathbf}

\makeatletter
\def\ps@IEEEtitlepagestyle{%
	\def\@oddfoot{\mycopyrightnotice}%
	\def\@evenfoot{}%
}
\def\mycopyrightnotice{%
	{\footnotesize \begin{minipage}{1\textwidth}
		© 2020 IEEE.  Personal use of this material is permitted.  Permission from IEEE must be obtained for all other uses, in any current or future media, including \newline reprinting/republishing this material for advertising or promotional purposes, creating new collective works, for resale or redistribution to servers or lists, or reuse of any copyrighted component of this work in other works.\hfill
		\end{minipage}}
	\gdef\mycopyrightnotice{}
}

	\def\@oddhead{\footnotesize \begin{minipage}{1\textwidth}
			Accepted at IEEE Transactions on Medical Imaging. Final version is available at  \url{http://dx.doi.org/10.1109/TMI.2020.3002695}\hfill \thepage
	\end{minipage}}%
	\def\@evenhead{}%

\begin{document}
\title{Appearance Learning for Image-based Motion Estimation in Tomography}

\author{Alexander~Preuhs,
	Michael~Manhart,
	Philipp~Roser, 
	Elisabeth~Hoppe,
	Yixing~Huang,
	Marios~Psychogios,
	Markus~Kowarschik,
	and~Andreas~Maier,~\IEEEmembership{Member,~IEEE}
	\thanks{Copyright (c) 2019 IEEE. Personal use of this material is permitted. However, permission to use this material for any other purposes must be obtained from the IEEE by sending a request to pubs-permissions@ieee.org.
		A.~Preuhs, P.~Roser, E.~Hoppe, Y.~Huang and A.~Maier are with the Pattern Recognition Lab, Friedrich-Alexander-Universität Erlangen-Nürnberg, 91058 Erlangen, Germany e-mail: alexander.preuhs@fau.de.}
	\thanks{M.~Psychogios is with the Neuroradiology Department, Universitätsspital Basel, 4031 Basel, Switzerland.}
	\thanks{M.~Manhart and M.~Kowarschik are with Siemens Healthcare GmbH, 91301 Forchheim, Germany.}}

%

\markboth{IEEE TRANSACTIONS ON MEDICAL IMAGING}%
{Preuhs \MakeLowercase{\textit{et al.}}: Appearance Learning for Image-based Motion Estimation}

\maketitle

\def\customlen{\hspace{0.472em}}

\def\cltwo{\hspace{0.555em}}

\def\meas{\vec{y}}
\def\atten{\vec{x}}
\def\sysmat{\vec{A}}
\def\proj{\vec{Y}}
\def\pm{\vec{P}}
\def\worldpoint{\vec{a}}
\def\featurenet{\mathcal{S}}
\def\regressionnet{\mathcal{R}}
\def\efftraj{\vec{E}}
\def\anatomicplane{\vec{V}}
\def\cnn{Cnn}
\def\ent{Ent}
\def\tv{Tv}
\def\cnnent{Cnn+}
\def\noco{None}
\def\groundtruth{Gt}
\def\degmm{$^\circ$\textbackslash mm}

\begin{abstract}
	In tomographic imaging, anatomical structures are reconstructed by applying a pseudo-inverse forward model to acquired signals.
	Geometric information within this process is usually depending on the system setting only, i.\,e., the scanner position or readout direction. 
	Patient motion therefore corrupts the geometry alignment in the reconstruction process resulting in motion artifacts. 
	We propose an appearance learning approach recognizing the structures of rigid motion independently from the scanned object. 
	To this end, we train a siamese triplet network to predict the \textit{reprojection error} (RPE) for the complete acquisition as well as an approximate distribution of the RPE along the single views from the reconstructed volume in a multi-task learning approach. 
	The RPE measures the motion-induced geometric deviations independent of the object based on virtual marker positions, which are available during training. 
	We train our network using 27 patients and deploy a 21-4-2 split for training, validation and testing. 
	In average, we achieve a residual mean RPE of 0.013\,mm with an inter-patient standard deviation of 0.022\,mm. 
	This is twice the accuracy compared to previously published results. 
	In a motion estimation benchmark the proposed approach achieves superior results in comparison with two state-of-the-art measures in nine out of twelve experiments. 
	The clinical applicability of the proposed method is demonstrated on a motion-affected clinical dataset.
\end{abstract}


\IEEEpeerreviewmaketitle
\section{Introduction}
\label{sec:introduction}
\IEEEPARstart{A}{ppearance} modeling \cite{Cootes1998} for interpreting images is a well examined problem in the field of computer vision. 
An appearance model is trained to extract invariant representations of an object of interest \cite{comaniciu2003kernel,grabner2006real}, e.\,g., for the tracking of faces \cite{babenko2010robust} or event detection \cite{xu2015learning}.
Recently, Preuhs et\,al.\,\cite{preuhs2019image} have applied the strategy of appearance learning for motion detection in tomographic imaging.  

The key concept of tomographic imaging is the reconstruction of internal patient anatomy from a series of measured signals. 
This can be the relaxation properties of hydrogen atoms in \textit{magnetic resonance imaging} (MRI)  or photon attenuation in X-ray \textit{computed tomography} (CT). 
When reconstructing a tomographic image from measured signals, the geometry associated with each signal only depends on the system setting, i.\,e., source-detector orientation of a CT system or readout position for MRI scanners. 
The object itself is assumed to be static during the  acquisition. 
As a consequence, patient motion corrupts the geometry alignment and results in motion artifacts within the reconstructed tomographic image.

Many efforts have been devoted to the problem of non-static objects, which are mainly splitted into  non-rigid  and  rigid approaches. 
Rigid approaches reduce the number of unknowns to a 6 dimensional vector per measured signal, i.\,e., the respective rigid patient pose. 
However, complex movements, as apparent in heart imaging, are not reducible to such a simple model. 
In these cases non-rigid motion estimation must be deployed.

\subsection{Non-Rigid Motion Compensation}
\label{subsec:nonrigidmotioncompensation}
Lauritsch et\,al.\,\cite{lauritsch2006towards} presented a gating approach, where the signal is binned to different motion states. 
Only similar motion states are used for reconstruction. 
This is extended by Taubmann et\,al.\,\cite{taubmann2016convex} who developed a primal-dual optimization scheme based on a spatial and temporal \textit{total variation} (TV). 
Gating approaches were also presented by Larson et\,al.\,\cite{larson2004self}  and Hoppe et\,al.\,\cite{hoppefree} for cardiac cine MRI, where the motion bin is deduced from the k-space center of each readout.
Similar to gating, Fischer et\,al.\,\cite{fischer2017mr} devised an MRI-based model for X-ray fluoroscopy overlays. 
By binning of 4-D volumes to cardiac and respiratory phases, the motion field is estimated using 3-D/3-D registration. 

Recent approaches deploy image-to-image translation from motion-affected  reconstructions to such without motion artifacts.  
Here, prior knowledge on the expected manifold of motion free reconstructions is learned \cite{zhu2018image}. 
Kustner et\,al.\,\cite{kustner2019retrospective} and Latif et\,al.\,\cite{latif2018automating} propose a \textit{conditional generative adversarial network} (cGAN) to synthesize motion free MRI reconstructions from a motion degenerated one. 
The same approach was presented for X-ray imaging by Xiao et\,al.\,\cite{xiao2019x}.

\subsection{Rigid Motion Compensation}
\label{subsec:rigidmotioncompensation}
For many anatomical objects, the structure of the expected motion is already known a priori. 
The head, for example, is restricted by the skull to move as a rigid object. 
Further, many anatomies move in an approximate rigid structure during interventions, e.\,g., the knees or the hands. 
As the focus of this article is rigid motion compensation, we give a detailed overview of published methods which can be clustered into three categories. 
\subsubsection{Projection Consistency}
\label{subsubsec:projectionconsistency}
A computationally fast approach is projection consistency, where only the projection raw data are used, without the need for intermediate reconstructions. 
The main idea is that information is redundantly sampled by the forward operator with each acquired signal, e.\,g., the mass of the object. 
Powerful conditions are the \textit{Helgason-Ludwig consistency conditions} (HLCC) \cite{helgasonradon} describing the relation between polynomials of degree $n$ and the respective $n$\textsuperscript{th} moment of the projections for parallel-beam CT or radial sampled MR. 
This was devised by Yu et\,al.\,\cite{yu2007data} to compensate motion in fan-beam geometry. 
A more broadly applicable approach based on the zero order HLCC and Grangeat's theorem is epipolar consistency which was applied for geometric jitter and motion compensation in \textit{cone-beam computed tomography} (CBCT) \cite{preuhs2019symmetry,preuhs2018double,Maass2014,Frysch2015}. 
Similar approaches have been deployed in MRI motion compensation, where propeller  trajectories measure the k-space center redundantly and compensate the motion based on this data redundancy \cite{pipe1999motion}.

\subsubsection{Reconstruction Consistency}
\label{subsubsec:reconstructionconsistency}
Contrary to projection consistency, reconstruction consistency solely uses tomographic images to estimate a rigid motion trajectory and is therefore often related as autofocus. 
The key idea is similar to the image-to-image translation approaches presented above: a motion-free reconstruction reveals some inherent properties which can be measured using an \textit{image-quality metric} (IQM). 
In contrast to cGAN-based approaches, however, a motion trajectory is estimated by iterative optimization of the IQM. 
This ensures data integrity, which is of high importance in a clinical setting. 
The first application of this autofocus principle was presented by Atkinson et\,al.\,\cite{atkinson1997autofocus} for MRI reconstructions. 
They optimize a motion trajectory to find a reconstructed image with low entropy of the intensity histogram favoring images with high contrast structures and without motion ghosting or blur. 
Kingston et\,al.\,\cite{Kingston2011} presented a similar approach based on TV minimization. 
Subsequently various extensions were proposed \cite{wicklein2012image,rohkohl2013improving,sisniega2017motion,herbst2019misalignment}, including a combination of metrics as well as additional smoothness constraints.

\subsubsection{Data Consistency}
\label{subsubsec:dataconsisntency}
The last category is based on enforcing data fidelity, which is the consistency of the reconstruction domain with the signal domain.
In CBCT this approach is used for calibrating the system geometry by minimizing a \textit{reprojection error} (RPE) of 3-D spheres on a calibration phantom and their respective 2-D projections  \cite{strobel2003improving}. 
Using markers attached to the patient, this strategy was also investigated for motion compensation in MRI \cite{ooi2009prospective} and CT \cite{muller2015fully}. 

A second approach to enforce data fidelity is the virtual application of the forward model to an intermediate reconstruction and comparing these virtual data with the actually acquired data. 
Haskell et\,al.\,\cite{haskell2018targeted} used a SENSE forward model to maximize the data consistency with the acquired k-space data. 
For transmission imaging, \textit{digitally rendered radiographs} (DRR) are commonly used to enforce consistency with the acquired projections $\cite{berger2016marker,Ouadah2016}$. 
In this context,
Dennerlein et\,al.\,\cite{Dennerlein2012} exploit directly and indirectly filtered projections to compensate for geometric misalignment.

\subsection{Potentials and Limitations in the State of the Art}
\label{subsec:limitationsandprotentials}
Non-rigid approaches (see\,Sec\,\ref{subsec:nonrigidmotioncompensation}) seem to be unfitting if the problem is known to be rigid.
Image-to-image-based approaches do not exploit the full problem knowledge. 
Furthermore, their clinical applicability is limited because the consistency of the reconstructed image to the acquired data is not guaranteed and anatomic malformations can vanish \cite{huang2019data}.

Consistency conditions (see\,Sec.\,\ref{subsubsec:projectionconsistency}) have been used for the compensation of various other image artifacts as beam hardening, scatter correction or truncation correction \cite{abdurahman2018beam,hoffmannempirical,wurfl2017epipolar,punzet2018extrapolation}.
This is due to consistency being deduced from a physical model, which only holds on an approximate basis for real applications \cite{preuhs2019maximum}. 
Additionally, they are insensitive to certain motion directions and their application is limited to motion patterns outside the acquisition plane \cite{preuhs2019symmetry,Frysch2015}.

Image-based methods (see\,Sec.\,\ref{subsubsec:reconstructionconsistency}) currently use hand-crafted features not particularly designed for the specific task. 
As a consequence, they are object dependent with each object revealing a different histogram entropy or TV.

A robust approach is based on reducing the RPE using markers (see\,Sec.\,\ref{subsubsec:dataconsisntency}). 
However, this approach depends on  additional marker placements, which has not found its way to clinical routine yet.  
Marker-free registration approaches are only working robustly if a prior reconstruction is available. 
Otherwise, the optimization becomes ill-posed, as the intermediate reconstruction, on which the forward model is applied, inherently reveals motion artifacts.

Deep learning has high potential to overcome some of those limitations by replacing bottlenecks of traditional methods with data-driven algorithms. 
For example, Bier et\,al.\,\cite{bier2018detecting} tackled the problem of manual marker placement by learning anatomical landmarks directly from the projection images. 
The presented cGAN-based approaches potentially have the risk of vanishing anatomical malformations, however, they may solve the chicken-egg problem for marker-free registration approaches.
Additionally, many applications emerged for learning-based registration \cite{liao2019multiview,toth20183d,chou20132d}. 
They could potentially be extended for motion compensation scenarios.

\subsection{Contribution}
\label{subsec:contribution}
Despite its great potential in improving rigid motion compensation algorithms, deep learning methods have caught limited attention from the research community. 
In Preuhs et\,al.\,\cite{preuhs2019image}, we have presented the concept of learning image artifacts from a single axial reconstructed slice using a simplified motion model and a vanilla network architecture. 
The key concept is that a certain motion state is regressed to an object-independent measure defined by the RPE. 
We extend this line of thinking by developing a new data-driven approach for appearance learning capable of compensating motion artifacts. 
Our network architecture for motion appearance learning is based on a siamese triplet network trained in a multi-task scenario. 
Therefore, we incorporate not only a single axial slice but make use of information from 9 slices, extracted from axial, sagittal and coronal orientations. 
Using a multi-task loss, we estimate both (1) an overall motion score of the reconstructed volume similar to \cite{preuhs2019image} and (2) a prediction which projections are affected by the motion.
To stabilize the network prediction, we deploy a novel pre-processing scheme to compensate for training data variability. 
These extensions allow us to learn realistic motion appearance, composed of three translation and three rotation parameters per acquired view. 
We evaluate the accuracy of the motion appearance learning in dependence of the patient anatomy and also the motion type.
In a rigid motion estimation benchmark, we demonstrate the performance of the appearance learning approach in comparison to state-of-the-art methods. Finally, we demonstrate its applicability to real clinical data using a motion-affected clinical scan.

We devise the proposed framework for CBCT, however, by exchanging the backward model and training data, this approach is seamlessly applicable to radial sampled MRI or \textit{positron emission tomography} (PET). 
In addition, by adjusting the regression target, also for Cartesian sampled MRI.

\section{Rigid Motion Model for CBCT}
\subsection{Cone-Beam Reconstruction}
\label{subsec:conebeamreconstruction}
In tomographic reconstruction we compute anatomical structures denoted by $\atten$ from measurements $\meas$ produced with a forward model $\sysmat$ by $\sysmat\atten=\meas$. 
For X-ray transmission imaging $\atten$ are attenuation coefficients and $\meas$ are the attenuation line integrals measured at each detector pixel. 
The system geometry~---~e.\,g., pixel spacing, detector size and source-detector  orientation~---~is part of the forward model $\sysmat$. 
Using the pseudo-inverse
\begin{equation}
\atten = \sysmat^\top (\sysmat\sysmat^\top)^{-1} \meas
\end{equation}
we get an analytic solution to this inverse problem, which consists of the back-projection $\sysmat^\top$ of filtered projection data $(\sysmat\sysmat^\top)^{-1} \meas$ \cite{maier2019learning}, commonly known as \textit{filtered back-projection} (FBP). 
For CBCT with circular trajectories, an approximate solution is provided by the \textit{Feldkamp-Davis-Kress} (FDK) algorithm  \cite{feldkamp1984practical}. 
The algorithm is regularly used for autofocus approaches \cite{Kingston2011,sisniega2017motion} (see\,Sec.\,\ref{subsubsec:reconstructionconsistency}) due to its low computational costs. 
Rit\,et\,al.\,\cite{rit2009comparison} have further shown that even due to its approximate nature, an FDK-based motion-compensated CBCT reconstruction is capable of correcting most motion artifacts. 
Thus, we use the FDK reconstruction algorithm, having the benefit of only filtering the projection images once and thereafter only altering the back-projection operator for motion trajectory estimation. 

\begin{figure}
	\centering
	\includegraphics[width=\linewidth]{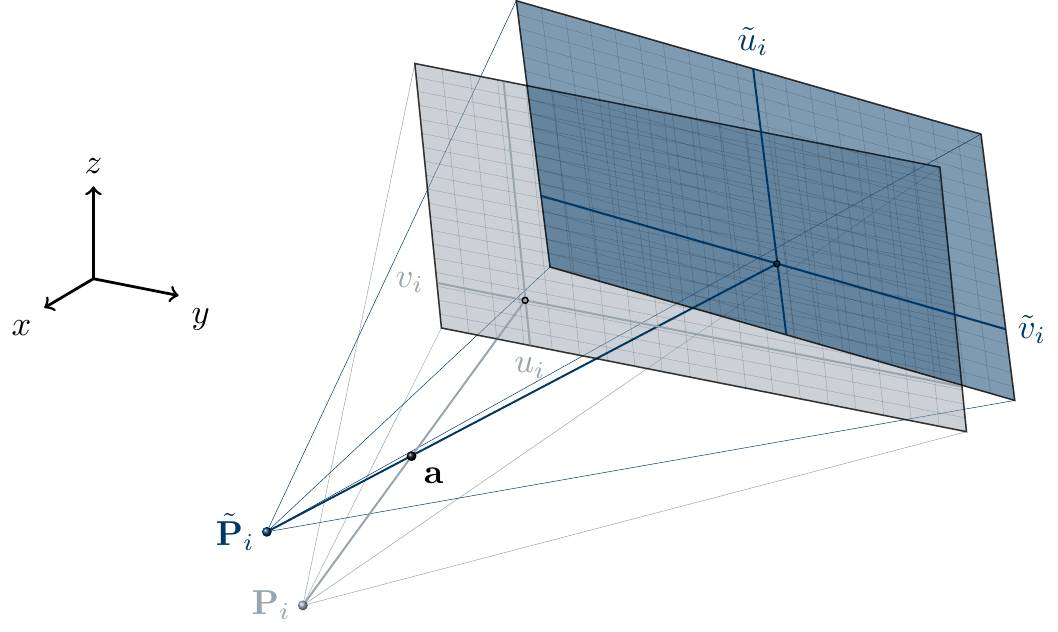}
	\caption{Visualization of the geometry for a point $\vec{a}$ and two geometries $\vec{P}_i$ and $\tilde{\vec{P}}_i$. The $L_2$ distance between the two projected points on the 2-D detector defines the RPE of the scene.}
	\label{fig:rpevisualization}
\end{figure}

It is possible to formulate the FDK algorithm using a tuple of projection matrices $\pm = (\pm_0, \pm_1, ...,\pm_N)$ describing the geometry of operator $\sysmat$. The measurements $\meas$ are reshaped to a tuple of \mbox{2-D} projection images $\proj = (\proj_0, \proj_0, ...,\proj_N)$.
In analogy to \cite{feldkamp1984practical}, we implement the FDK for a short scan trajectory using Parker redundancy weights $W_i(u,v)$ \cite{parker1982optimal}, where $i \in [1,2, ..., N]$ describes the projection index and $(u,v)$ denotes a 2-D pixel. 
The first step is a weighting and filtering of the projection images
\begin{equation}
\proj^\prime_i (u,v) = W_i(u,v) \int_{\mathbb{R}} \mathcal{F} \, \tilde{\proj}_i(\eta,v) \, \textnormal{e}^{i 2 \pi u v} \frac{|\eta|}{2} \, \textnormal{d} \eta \enspace,
\end{equation} 
with $\mathcal{F} \, \tilde{\proj}_i$ being the 1-D Fourier transform of the $i$\textsuperscript{th} cosine weighted projection image along the tangential direction of the scan orbit.  
Thereafter, a distance-weighted voxel-based back-projection is applied  mapping a homogeneous world point $\worldpoint \in \mathbb{P}^3$ to a detector pixel described in the projective two-space $\mathbb{P}^2$
\begin{equation}
f_\textnormal{FDK}(\worldpoint,\pm,\proj) = \sum_{i \in N} U(\vec{P}_i,\worldpoint) \proj^\prime_i(\phi_u(\pm_i \worldpoint), \phi_v(\pm_i \worldpoint))
\label{eq:fdk}
\end{equation}
with $\pm_i$ describing the system calibration associated with $\proj_i$. (see\,Fig.\,\ref{fig:rpevisualization}). 
The mapping function $\phi_\diamond : \mathbb{P}^2 \rightarrow \mathbb{R}$ is a dehomogenization
\begin{equation}
\phi_\diamond((x,y,w)^\top) = 
\begin{cases}
\frac{x}{w} & \textnormal{if $\diamond = u$} \\
\frac{y}{w} & \textnormal{if $\diamond = v$} 
\end{cases} \enspace,
\label{eq:dehomogenization}
\end{equation}
and $U(\vec{P}_i,\worldpoint)$ is the distance weighting according to \cite{feldkamp1984practical}. 

\subsection{Rigid Motion Model}
\label{subsec:rigidmotionmodel}

We assume the rigid motion to be discrete w.\,r.\,t.~the acquired projections. 
To this end, we define the motion trajectory $\vec{M}$ as a tuple of motion states $\vec{M}_i \in \mathbb{SE}(3)$ describing the orientation of the patient during the acquisition of the $i$\textsuperscript{th} projection $\proj_i$.
Each motion state is associated to a projection matrix $\vec{P}_i$. The motion modulated trajectory is obtained by 
\begin{equation}
\vec{P} \circ \vec{M} = (\vec{P}_0\vec{M}_0, \, \vec{P}_1\vec{M}_1, \, ..., \, \vec{P}_N\vec{M}_N) \enspace,
\end{equation}
where $\circ$ is the element-wise matrix multiplication of two tuples. 
Typically, the motion trajectory is unknown and the task of motion compensation is to find a tuple of matrices $\vec{C}_i \in \mathbb{SE}(3)$ annihilating the resulting geometry corruption produced by $\vec{M}$. 
The compensation is successful if an annihilating trajectory $\vec{C} = (\vec{C}_0, ...\vec{C}_N)$ is found that suffices  $\vec{C} \circ \vec{M} = \vec{1}$, with $\vec{1}$ being a tuple of identities.

Each motion matrix defined in $\mathbb{SE}(3)$ is parameterized by 3 rotations ($r_x,r_y,r_z$) and 3 translations ($t_x,t_y,t_z)$, describing Euler angles and translations along  the corresponding coordinate axis, respectively.
Therefore, the annihilating trajectory has $6N$ free parameters for an acquisition with $N$ projections.
To reduce the high dimensionality, we model the trajectory using Akima splines \cite{akima1970new}. This reduces the free parameters to $6M$, where $M$ is the number of nodes typically chosen as $M \ll N$. Based on the expected frequency of the motion the number of spline nodes can be adapted.

\section{Appearance Learning}
Conventionally, autofocus approaches are based on hand-crafted features, selected due to their correlation with an artifact-free reconstruction. 
For example, entropy gives a measure on contingency. As the human anatomy consists of mostly homogeneous tissues, entropy of the gray-value histogram can be expected to be minimal if all structures are reconstructed correctly.  
Motion blurs the anatomy or produces ghosting artifacts distributing the gray values more randomly.
A similar rational is arguable for TV, which is also regularly used for constraining algebraic reconstruction \cite{taubmann2016convex}. 
Contrary to algebraic reconstruction, the motion estimation scenario is non-convex  and optimization of a cost function based on hand-crafted image features is hardly solvable for geometric deviations  exceeding a certain bound \cite{herbst2019misalignment}. 

We aim to overcome this problem by designing a tailored image-based metric, which reflects the appearance of the motion structure independent of the object.
\subsection{Object-Independent Motion Measure}
Several metrics have been proposed to quantify image quality of motion affected reconstructions based on a given ground truth: the \textit{structural similarity} (SSIM) \cite{silvia2019towards}, the $L_2$ distance \cite{braun2018motion} or binary classification to motion-free and -affected \cite{meding2017automatic}. 
However, they were not used for the compensation of motion, but merely for the assessment of image quality, which is of high relevance in the field of MRI to automize prospective motion compensation techniques.

We choose the object-independent RPE for motion quantification. Its geometric interpretation  is schematically illustrated in Fig.\,\ref{fig:rpevisualization}. 
The RPE measures reconstruction relevant deviations in the projection geometry and is defined by a 3-D marker position $\worldpoint \in \mathbb{P}^3$ and two projection geometries $\pm_i, \tilde{\pm}_i$. 
We consider $\pm_i$ as the system calibration and $\tilde{\pm}_i = \pm_i \vec{M}_i$ as the actual geometry due to the patient motion. 
Accordingly, the RPE for a patient movement at projection $i$ is defined by
\begin{equation}
d_\textnormal{RPE}(\vec{P}_i,\tilde{\pm}_i,\worldpoint) =  \left\Vert \, \begin{pmatrix} 
\phi_u(\pm_i \worldpoint)         \\[0.3em]
\phi_v(\pm_i \worldpoint) \\[0.3em]
\end{pmatrix}
- 
\begin{pmatrix}
\phi_u({\tilde{\pm}_i} \worldpoint)        \\[0.3em]
\phi_v({\tilde{\pm}_i} \worldpoint) \\[0.3em]
\end{pmatrix}\, \right\Vert_2^2
\label{eq:projectionrpe}
\end{equation}
where $\phi_\diamond$ denotes the dehomogenization described in Eq.\,\eqref{eq:dehomogenization}.  Using a single marker, the RPE is insensitive to a variety of motion directions.
Therefore, we use $K=90$ virtual marker positions $\worldpoint_k$, distributed homogeneously at three sphere surfaces with the radii $30$\,mm, $60$\,mm, and $90$\,mm. 
The high number of markers ensures that the RPE is view-independent, i.\,e., a displacement of a projection at  the beginning of the trajectory has the same effect on the RPE as a displacement of a projection at the end of the trajectory.
Accordingly, the overall RPE for a single view is
\begin{equation}
d_\textnormal{RPE}(\vec{P}_i,\tilde{\vec{P}}_i) = \frac{1}{K} \sum_{k=1}^K d_\textnormal{RPE}(\vec{P}_i,\tilde{\vec{P}}_i,\worldpoint_k) \enspace.
\label{eq:rpe}
\end{equation}
As shown in Strobel et\,al. \cite{strobel2003improving}, Eq.\,\eqref{eq:rpe} can be rewritten to a measurement matrix $\vec{X}$ containing the \mbox{3-D} marker positions, a vector $\vec{p}$ containing the elements of $\tilde{\vec{P}}_i$ and a vector $\vec{d}$ containing the respective \mbox{2-D} marker positions. Given at least six markers, the components of $\vec{p}$ are estimated as the solution to $\Vert\vec{X} \vec{p} - \vec{d} \Vert^2_2$.  Direct application of this method for motion compensation is non-trivial, as the  accurate estimation of $\worldpoint_k$ is challenging. The 3-D marker positions must be estimated from projection images with corrupted geometry alignment.

Thus, we follow a different approach: we train a neural network to predict the RPE directly from the reconstructed images. To generate training data, we simulate rigid motion on real clinical acquisitions and compute the corresponding ground truth RPE via the virtual marker positions and their corresponding projections using Eq.\,\eqref{eq:rpe}.
Thus, we aim to approximate Eq.\,\eqref{eq:rpe} from reconstructed slice images using a neural network. 

\subsection{Network Architecture}
\label{subsec:networkarchitectur}
\begin{figure*}
	\includegraphics[width=\linewidth]{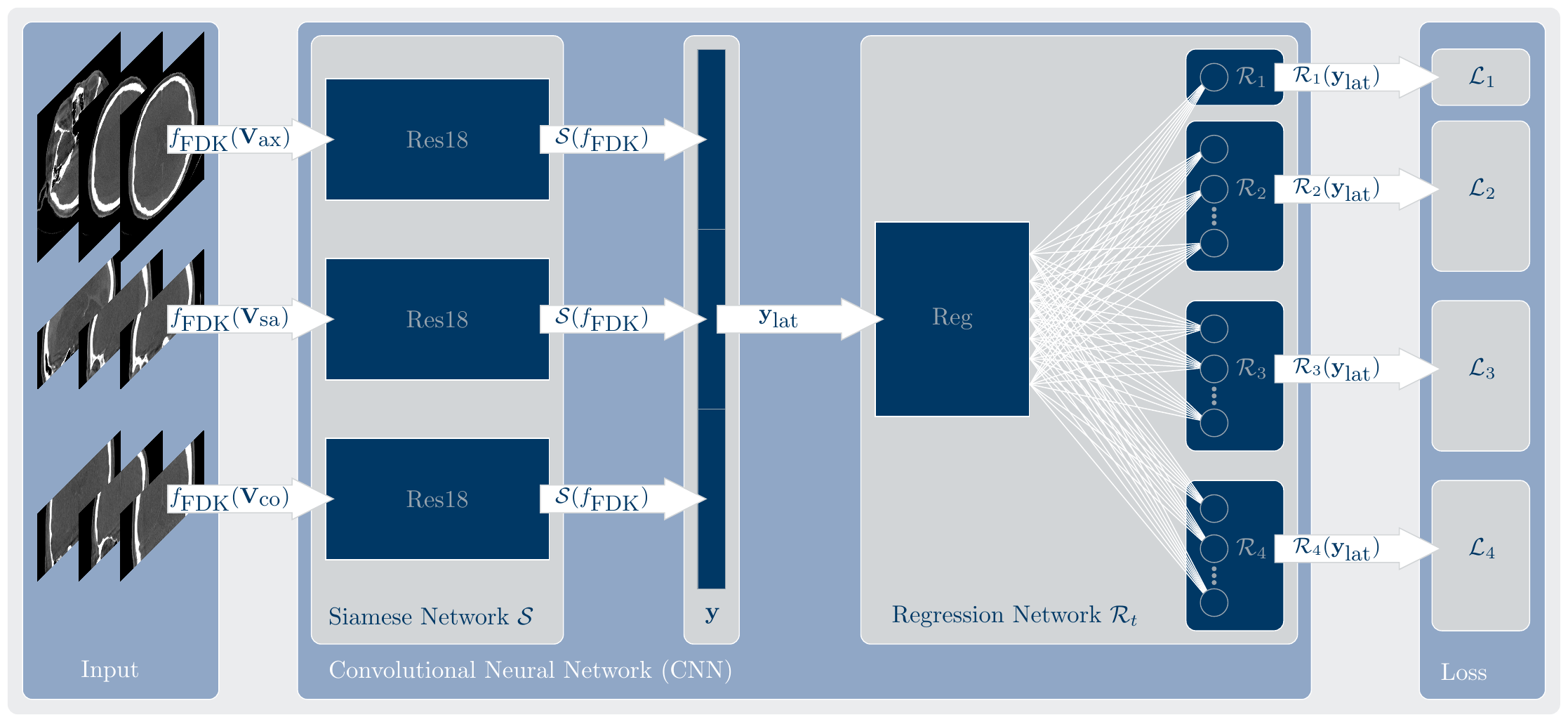}
	\caption{Flowchart of network architecture. The input to the siamese triplet network are three slices of different anatomical orientations. The concatenated output is fed to the multi-task regression network. Based on the four outputs the respective loss is computed.}
	\label{fig:networkarchitecture}
\end{figure*}
Our network architecture depicted in Fig.\,\ref{fig:networkarchitecture} consists of two stages, a feature extraction stage followed by a regression stage. 
The feature extraction is driven by a siamese triplet network architecture consisting of three weight-sharing feed forward networks denoted by $\mathcal{\featurenet}$. The output of the three networks is concatenated and fed to the regression network $\regressionnet_t$. 
The feed forward network is almost identical to the ResNet-18 architecture \cite{he2016deep} upto the last global average pooling. 
We devise the network to our task by removing the last \textit{fully connected} (FC) layer. 
Since the input of our network is always a tomographic reconstruction, we also remove all \textit{batch normalization} (BN) layers. 
Expecting three-channel input images ranging from $\mathbb{R}^{70\times216}$ to $\mathbb{R}^{256\times216}$ for the different anatomical orientations, the final $7\times7$ average pooling is replaced by a $3\times3$ average pooling. 
The resulting feature maps are concatenated and represent the input to the regression network. 

The regression network $\regressionnet_t$ is composed of a $1\times1$ convolution mapping the $1536\times3\times3$ feature maps to $2048 \times3\times3$ feature maps followed by an $1\times1$ global average pooling. 
The resulting feature maps are fed to four FC layers, each representing a different task $t \in \{1,2,3,4\}$. The first FC layer maps to a single scalar output  $\regressionnet_1$, the other three FC layers map to $N$ dimensional outputs  $\regressionnet_2$, $\regressionnet_3$ and $\regressionnet_4$, where $N$ represents the number of projections.

\subsection{Data Generation} 
\label{subsec:datageneration}
Motion-affected reconstructions with corresponding ground truth motion patterns are rarely available. 
First, contrary to spiral CT, CBCT patient data are not available from public sources and therefore difficult to obtain in general. 
Second, the only robust motion compensation is based on external markers, which is not used in clinical practice. 
The only feasible possibility is the generation of artificial motion based on motion-free acquisitions. 
To this end, our data-base consists of 27 clinical head CBCTs, each being ensured to have no motion artifacts by a medical expert. 
The data are acquired with a clinical CBCT system (Artis Q, Siemens Healthcare GmbH, Forchheim, Germany). 
After filtering, the high resolution projection images are down-sampled to low resolution projection images $\proj_i \in \mathbb{R}^{320\times413}$ using an average filter. 
This improves the computational performance of the method and matches the voxel size of the volume reconstructed from these images. 
Down-sampling does barely affect the accuracy of the autofocus method \cite{atkinson1997autofocus}.

The first step of the data generation is an alignment process of the 27 clinical CBCT scans to a mean shape. 
We perform this semi-automatically based on a symmetry plane alignment \cite{preuhs2019symmetry}. The result of the alignment process is a single rigid transformation which is incorporated into the system trajectory $\vec{P}$.

The second step is the generation of rigid motion which can be realized by two approaches: (1) the reconstructed volume is reprojected on a motion modulated trajectory using DRRs and then reconstructed again using $\vec{P}$ or (2) the calibrated system trajectory $\vec{P}$ is virtually altered by a motion trajectory and reconstructed. 
As DRRs are simulated projections, they cannot model the complexity of a real system and alter resolution and noise  characteristics, where the latter is known to be critical for \textit{convolutional neural network}s (CNN) \cite{huang2019data,maier2019learning}. 
Therefore, we decide to choose strategy (2) where projection characteristics of a real clinical setting are preserved.
Further, note that $\mathbb{P}^3$ is diffeomorphic to $\mathbb{SE}(3)$, as a consequence a rigid motion can be analogously expressed by a transformation of the system geometry $(\vec{P}_i\vec{T}_i)\,\worldpoint$ or a transformation of the object $\vec{P}_i(\vec{T}_i \worldpoint)$. 

The motion generation is applied as follows: First, the system calibration $\vec{P}$ is altered by a motion trajectory $\vec{M}$, giving the effective trajectory $\vec{E} = \vec{P}\circ\vec{M}$. 
The motion trajectory contains a random misalignment in one of the 6 motion splines. 
The length of misaligned motions is chosen to be distributed over a third of the trajectory and restricted to views unaffected by the Parker redundancy weighting (see\,Sec.\,\ref{subsec:conebeamreconstruction}). 
The redundancy weighting alters the appearance of motion artifacts~---~e.\,g., any translations of the last few views would barely affect the reconstruction quality as those projections are mostly outfaded~---~making a consistent mapping of artifact pattern to RPE infeasible. 
Secondly, two values are computed, the corresponding RPE per view using Eq.\,\eqref{eq:rpe} and the motion affected reconstruction using Eq.\,\eqref{eq:fdk}. 
Note that the RPE is only computed based on virtual marker positions which is possible because we know the system calibration $\vec{P}$ and motion trajectory $\vec{M}$ during training.
The volume is reconstructed on nine slices distributed in triplets of axial ($\mathbb{R}^{216\times256}$), coronal ($\mathbb{R}^{70\times216}$) and sagittal ($\mathbb{R}^{70\times256}$) slices with an isotropic voxel size of $0.84$\,mm. 
The respective slices are distributed in a volume $(\mathbb{R}^{256\times216\times70})$. 
The field-of-view is selected such that no truncation artifacts in longitudinal direction are present if reconstructed from a typical clinical scan.

\subsection{Motion Learning}
\label{subsec:motionlearning}
\def\reco#1{{\boldsymbol\mu}_\textnormal{FDK}^{#1}(\vec{E})}

The overall goal of the motion learning process is to train a network that is capable of approximating Eq.\,\eqref{eq:rpe} only based on a tomographic reconstruction. 
Therefore, nine slices of the tomographic reconstruction are used as input to the network. 
To keep the computational effort on a minimum and still capture all types of motion artifacts, the input of the network are triplets of three slices oriented in \textit{axial} (ax), \textit{coronal} (co) and \textit{sagittal} (sa) direction. 
We denote the respective coordinates of the nine slices by $\vec{V} = (\vec{V}_\textnormal{ax},\vec{V}_\textnormal{co},\vec{V}_\textnormal{sa})$, where $\vec{V}_\diamond$ is a set of coordinates defined in $\mathbb{P}^3$. 
Thus, $f_\textnormal{FDK}(\vec{V}_\textnormal{ax},\vec{E},\proj)$ will denote the reconstruction of three slices in axial direction with effective trajectory $\vec{E}$. 
Let us define a triplet of reconstructed slice images for a set of projections $\proj$ reconstructed with the effective trajectory $\vec{E}$ as $\reco{\diamond} = f_\textnormal{FDK}(\vec{V}_\diamond,\vec{E},\proj)$ and further, let the tuple of all reconstructed slices be $\reco{} = (\reco{\textnormal{ax}}, \reco{\textnormal{co}}, \reco{\textnormal{sa}})$. Then,  the input to the regression network is computed as 
\begin{equation}
\vec{y}_\textnormal{lat}(\reco{}) = \cup_{\diamond \in {\{\textnormal{ax},\textnormal{co},\textnormal{sa}\}}} \featurenet(\reco{\diamond}) \enspace,
\end{equation}
where $\cup$ denotes concatenation. Thus, each feed forward network processes three slices of the same anatomical orientation and the result is concatenated representing the latent space %
$\vec{y}_\textnormal{lat}(\reco{})$.
The loss function $l$ is based on a multi-task loss
\begin{equation}
l(\reco{}, \vec{E}) = 
\sum_{t=1}^4  
\Vert 
\regressionnet_t(\vec{y}_\textnormal{lat}(\reco{}) )
\\ - \mathcal{L}_t(\efftraj) 
\Vert_2^2 \enspace,
\end{equation}
with
\begin{equation}
\mathcal{L}_t(\efftraj) = 
\begin{cases}
\frac{1}{N} \sum_{i=1}^N d_\textnormal{RPE}(\efftraj_i) & \textnormal{if $t = 1$} \\
( d_\textnormal{RPE}(\efftraj_1),...,d_\textnormal{RPE}(\efftraj_N)) & \textnormal{if $t = 2$} \\
( d_\textnormal{RPE}(\efftraj^{\textnormal{ip}}_1),...,d_\textnormal{RPE}(\efftraj^{\textnormal{ip}}_N)) & \textnormal{if $t = 3$} \\
( d_\textnormal{RPE}(\efftraj^{\textnormal{op}}_1),...,d_\textnormal{RPE}(\efftraj^{\textnormal{op}}_N)) & \textnormal{if $t = 4$} \\
\end{cases} \enspace.
\end{equation}
Here, we assume that $\efftraj$ is implemented such that it can be decomposed into $\vec{P}$ and $\vec{M}$ allowing to compute the RPE using Eq.\,\eqref{eq:rpe}. 
$\efftraj^{\textnormal{op}}$ and $\efftraj^{\textnormal{ip}}$ refer to in-plane an out-plane motion. 
Assuming the system is rotating around the $\vec{z}$ axis, in-plane motion is within the acquisition plane and represented by parameters ($r_z,t_x,t_y$) and out-plane motion is stepping out the acquisition plane and represented by parameters ($r_y,r_z,t_z$). 
We use this distinction, because in-plane motion is better detectable in axial slices, whereas out-plane motion is better detectable in coronal and sagittal slices. 

For optimization we use the ADAM optimizer with a learning rate of $10^{-4}$ and a batch size of 32. 
To avoid over-fitting, we use the validation set for early stopping. 
The residual network $\mathcal{S}$ is initialized using pre-trained weights learned for the ImageNet classification task. 
The regression network $\mathcal{R}_k$ is randomly initialized.

\section{Experiments and Results}
In this section we evaluate the network performance w.\,r.\,t.\ 
three aspects: (1) the behavior of the network in its core task, i.\,e., the regression of the RPE, (2) the performance of the network in a motion compensation benchmark in comparison to state-of-the-art methods, and (3) the applicability of the proposed method to motion-affected clinical data.  
\subsection{Network Accuracy}
\label{subsec:networkaccuracy}
Using the data generation proposed in Sec.\,\ref{subsec:datageneration}, we generate 9001 different motion affected reconstructions. The amplitude of the applied motion is in the range of $0$ $^\circ$\textbackslash{}mm to $15$ $^\circ$\textbackslash{}mm, i.\,e.,  \textit{mean RPE}s (mRPE) are in a range of  $0$\,mm to $0.74$\,mm.
Using a \mbox{21-4-2} split, this provides us with a total of 189021 samples for training, 36004 samples for validation and 18002 samples for testing. 
The number of spline nodes is set to $M=20$. 
Following the training described in Sec.\,\ref{subsec:motionlearning}, we achieve the optimal validation loss after $12\times10^3$  iterations (see\,Tab.\,\ref{tab:networkperformance}).
\subsubsection{Ablation Study}
\begin{table}
	\caption{Best performing validation loss for different network settings.}
	\centering
	\label{tab:networkperformance}
	{\def\arraystretch{1.1}\tabcolsep=1pt
		\begin{tabular}{@{\extracolsep{10pt}} l c c c   c  @{}}\hline
			& $\mathcal{L}_1$ & $\mathcal{L}_2$ & $\mathcal{L}_3$& $\mathcal{L}_4$ \\ 
			\hline
			Proposed			&   \textbf{0.0098}  &0.2493 &  \textbf{ 0.1615} &\textbf{0.1981}   \\
			Proposed with BN 			&   0.0138  & 0.2835   & 0.2145 & 0.2412      \\
			Proposed~no alignment	 	&   0.0436  & 0.5644   &  0.5815 & 0.8192      \\
			Proposed~no pre-training  &   0.0171  & 0.4350	& 0.2941  &  0.3279	        \\
			Proposed~dual task  	    &   0.0146  &  \textbf{ 0.2481 }  &  x 		& x   \\
			Proposed with DenseNet 		    	&   0.0205  & 0.3331   &   0.1929 & 0.2818           \\ 
			\hline 	
		\end{tabular}
	}
\end{table}
To inspect the network performance as well as the influence of the pre-processing, Tab.\,\ref{tab:networkperformance} displays the respective best validation loss values for alterations in the network structure or input data. 
The most important performance boost is obtained by the pre-processing step of aligning the respective reconstructions and slight improvements are obtained by removing the BN. 
Further, a pre-training of $\mathcal{S}$ on ImageNet increases the accuracy. 
Without the distinction of in-plane and out-plane motion (dual-task learning), the accuracy of $\mathcal{L}_2$ decreases slightly, however, the mRPE ($\mathcal{L}_1$)  accuracy increases by $\approx50$~\%.  
A replacement of the residual network architecture $\mathcal{S}$ with a pre-trained DenseNet \cite{huang2017densely} worsens the accuracy.

\subsubsection{Patient and Motion Variability}
\label{subsubsec:paitentandmotionvariability}
\begin{figure}
	\centering
\includegraphics[width=\linewidth]{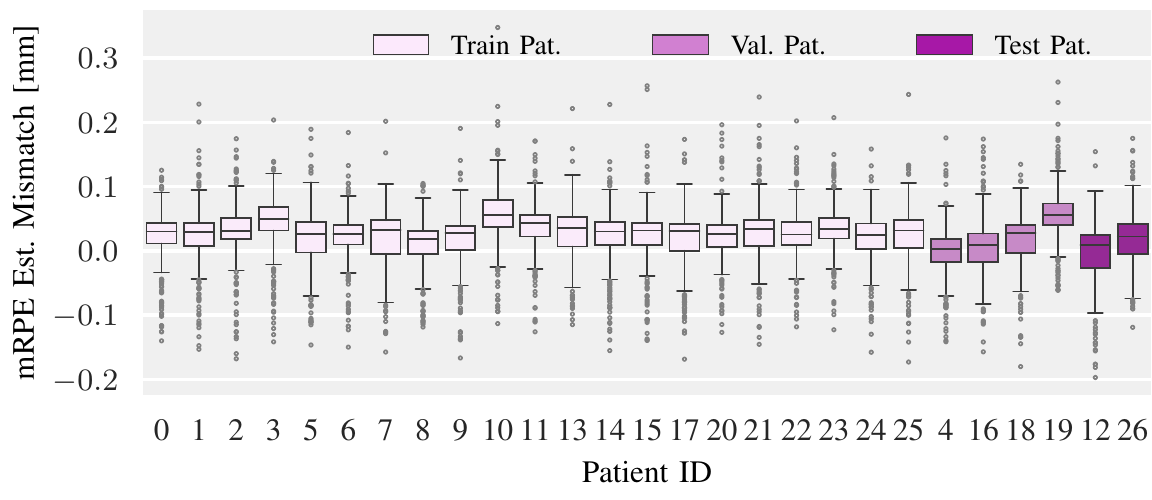}
	\caption{Boxplots showing the deviation of the mRPE ($\mathcal{L}_1$)  from the ground truth for each patient. The boxplots are grouped to training (Train Pat.), validation (Val. Pat.) and test (Test Pat.) patients. 
		All outliers  are displayed as circles.}
	\label{fig:boxplotmeanrpepatient}
\end{figure} 
An important aspect of motion appearance learning is the independence to the patient anatomy, similar motions applied to different patients should be predicted alike. 
Therefore, following the data generation presented in Sec.\,\ref{subsec:datageneration} we generate 300 additional motion shapes per patient ranging from mRPEs of $0$\,mm to $0.7$\,mm. 
Note, that the simulated motion is random and therefore not part of the training set. 
Consequently, the applied motion was never seen by the network. 
The results depicted in  Fig.\,\ref{fig:boxplotmeanrpepatient} show the patient-wise accuracy in predicting the mRPE ($\mathcal{L}_1$). 
Most of the outliers are within a range of $0.2$\,mm, and no outlier is exceeding an error of $0.35$\,mm. 
While the accuracy is high with an mRPE of 0.013\,mm, there is a slight tendency of overestimating the mRPE. 
The inter-patient variability of the estimation is small with a standard deviation of 0.022\,mm. 
From the patients never seen during training, we can observe a good generalization of the learned features. 
The tendency to overestimate the mRPE is even slightly less observable.  
\begin{figure}
	\centering
\includegraphics[width=\linewidth]{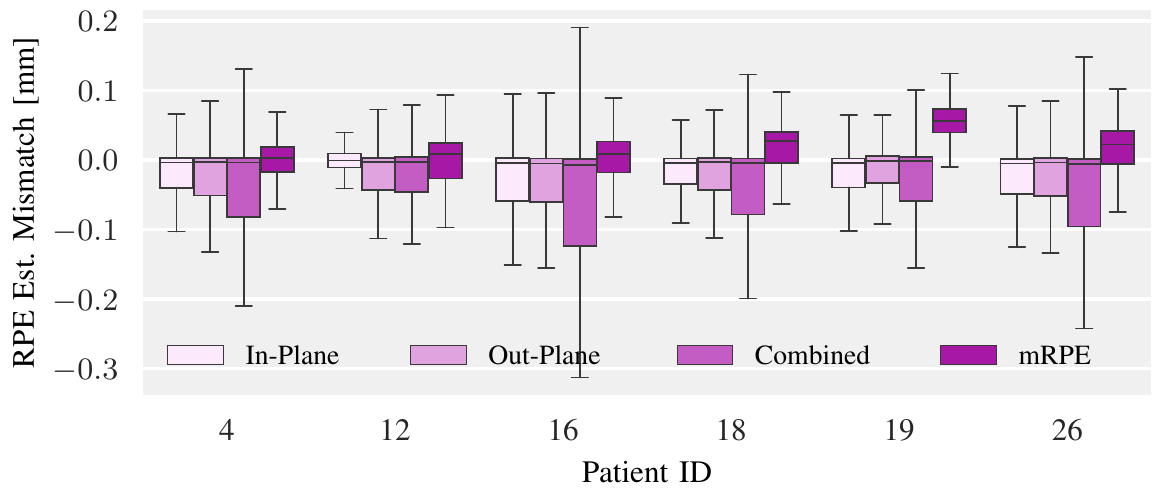}
	\caption{Boxplots visualizing the deviations to the ground truth for the in-plane ($\mathcal{L}_3$), out-plane ($\mathcal{L}_4$) and combined ($\mathcal{L}_2$) vRPE  and the mRPE ($\mathcal{L}_1$). 
		The evaluation is based on the four validation and two test patients.}
	\label{fig:boxplotrpeplots}
\end{figure} 
Besides the mRPE the network further predicts three \textit{view-wise RPE}s (vRPE) split to in-plane motion ($\mathcal{L}_2$), out-plane motion ($\mathcal{L}_3$) and both ($\mathcal{L}_4$). 
The accuracy for this task is depicted in Fig.\,\ref{fig:boxplotrpeplots}. 
Comparing the accuracy of the vRPE estimations to the mRPE we observe higher deviations and a higher number of outliers in the vRPE estimations. 
The accuracy of the in-plane vRPE is higher than for the out-plane vRPE. 
In-plane motion is mostly distributed in axial slices, which can be reconstructed without significant cone-beam artifacts and are best suitable for motion prediction. 
The patient-wise deviations are more pronounced compared to the mRPE, however, still on a reasonable low level.

\begin{figure}
	\centering
\includegraphics[width=\linewidth]{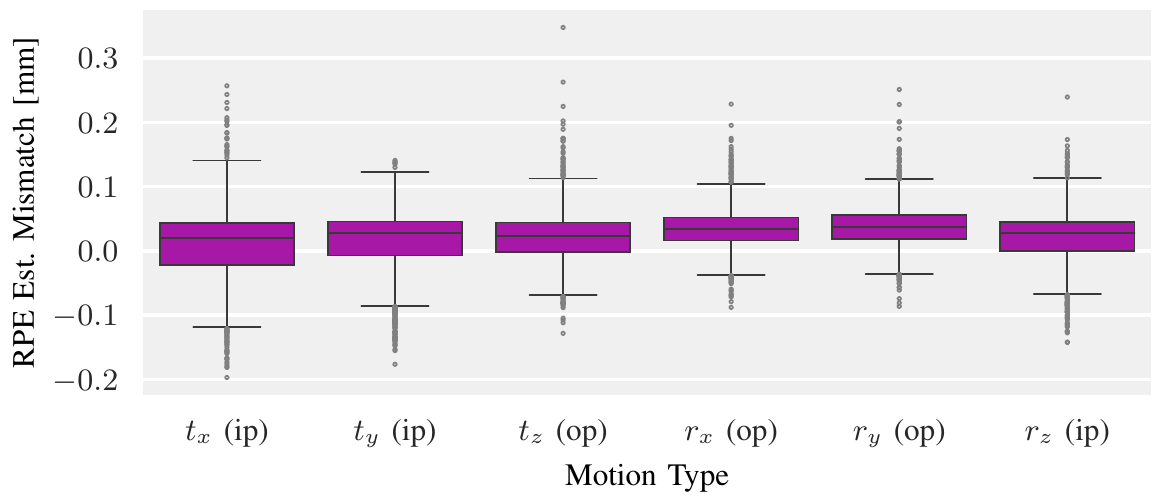}
	\caption{Boxplots showing the mRPE deviations from the ground truth for motions clustered to the 6 motion directions. 
		The evaluation is based on the four validation and two test patients. 
		All outliers  are displayed as circles.}
	\label{fig:boxplotmeanrpemovement}
\end{figure} 
Figure \ref{fig:boxplotmeanrpemovement} shows the mRPE clustered w.\,r.\,t.\ the motion directions for all patients. 
All motion directions can be predicted with similar accuracy, however, a slight tendency is observable that out-plane motion is predictable with less deviations.

In conclusion, the experiments have shown the patient independence of the proposed appearance learning approach, as well as the independence of the motion direction. 
This provides us a method that has no inherent limitations to certain  motion patterns as apparent, e.\,g., in epipolar consistency \cite{Frysch2015}, which is sensitive to out-plane motion, but barely applicable to in-plane motion. 
\subsubsection{RPE Trajectory Prediction}
\label{subsubsec:correlationstudy}
\begin{figure}
	\centering
\includegraphics[width=\linewidth]{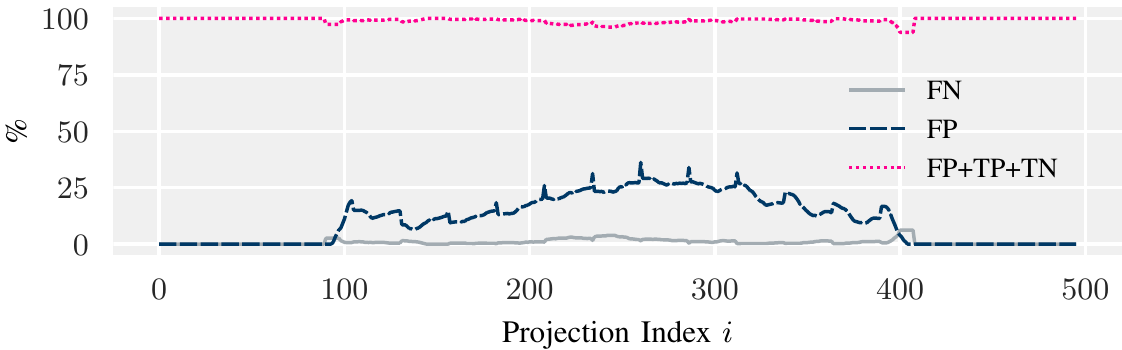}
	\caption{Binary soft-classification results plotted over the respective views. 
		FN relates to a motion-affected region that is classified by the network as motion-free. 
		FP relates to a misclassification to the motion-affected class. 
		TP and TN are correctly predicted views.}
	\label{fig:precision}
\end{figure}
Using the same data as in  Sec.\,\ref{subsubsec:paitentandmotionvariability}, we investigate in this experiment the performance of the estimated vRPE for motion classification. 
The predicted vRPEs are interpreted as soft-classifiers, where we define a view to be in the motion-free class (negative) if the average predicted value for view $i$ satisfies
\begin{equation}
\frac{1}{2} \mathcal{R}_2 + \frac{1}{4} \mathcal{R}_3 +\frac{1}{4} \mathcal{R}_4 \leq 0.1 \enspace.
\label{eq:softclassifier} 
\end{equation} 
In Fig.\,\ref{fig:precision} the accuracy is displayed encoded to \textit{false negative} (FN), \textit{false positive} (FP), and the combination of FP, \textit{true positive} (TP) and \textit{true negative} (TN). 
If the predicted value is used as an indicator function (see\,Sec.\ref{subsec:motionestimation}) in a motion estimation scenario a low FN rate is important. 
Regions classified as motion free will receive little attention within the optimization. 
On the opposite, a FP classification does not worsen the result and is therefore non-critical. 
These properties are satisfied as observable from Fig.\,\ref{fig:precision}. 
The FN rate is $\approx 0$\% and the FP rate is $\approx 25$\%. 
Note that the peaks in the FP curve are due to the spline nodes, where transitions from motion-affected to motion-free views  arise with increased frequency.

\subsection{Motion Estimation Benchmark}
\label{subsec:motionestimation}
\subsubsection{Autofocus}
The motion estimation benchmark is based on the four validation patients and two test patients. 
We apply a known motion trajectory $\vec{M}$ to the projection matrices $\vec{P}$ and evaluate the performance of six metrics (see\,\ref{subsubsec:imagequalitymetrics}) to find the annihilating trajectory $\vec{C}$ (see\,\ref{subsec:rigidmotionmodel}). 
We describe the trajectory as a function of six motion spline nodes $\vec{m} = (\vec{m}_{t_x},\vec{m}_{t_y},\vec{m}_{t_z},\vec{m}_{r_x},\vec{m}_{r_y},\vec{m}_{r_z})$. 
Each element of $\vec{m} \in \mathbb{R}^{6\times M}$ describes the respective spline node within the trajectory. 
Thus, $\vec{m}_{r_y,420}$ describes the rotation around the $y$-axis at acquisition view $420$. 
Then, the motion curve vector $\vec{t}(\vec{m}) = (\vec{t}_x,\vec{t}_y,\vec{t}_z,\vec{r}_x,\vec{r}_y,\vec{r}_z)$ is computed by evaluating the spline for each acquired view. 
For example $\vec{t}_x = (\eta_{\vec{m}_{t_x}}(0),\eta_{\vec{m}_{t_x}}(1), ...,\eta_{\vec{m}_{t_x}}(N))^\top$, with $\eta_{\vec{m}_{t_x}}(i)$ denoting the spline evaluation at position $i$ based on the spline nodes $\vec{m}_{t_x}$ as proposed in \cite{akima1970new}. 
From the six motion curves described by $\vec{t}$ we can directly compute the annihilating trajectories denoted by $\vec{C}(\vec{t}(\vec{m}))$. 
Note, that the motion trajectory itself is generated in an equal way.

The components of $\vec{m}$ are found by optimizing the IQM $f_\textnormal{IQM}$
\begin{equation}
\hat{\vec{m}} = \argmin_\vec{m} f_\textnormal{IQM}\left(f_\textnormal{FDK}(\vec{V}, \vec{P}\circ \vec{C}(\vec{m}), \proj)   \right) \enspace.
\label{eq:optimization}
\end{equation}

\subsubsection{Optimization}
\label{sec:optimization}
Equation \eqref{eq:optimization} is optimized using the gradient free downhill simplex algorithm \cite{olsson1975nelder}. 
We optimize only one node at a time iterating over all nodes in sequential order. 
We use a coarse to fine strategy by defining 5 stages. 
In the first three stages we define a starting stepsize of 1\,\degmm~for the simplex and set the number of iterations to 2. 
This allows a rough estimate of the trajectory. 
In the last two stages, we set the number of iterations to 100 with initial stepsize of 0.5\,\degmm. 
The optimization is finished if either the maximum number of iterations is exceeded or the improvement in $\vec{m}$ is below 0.001 \degmm.
\subsubsection{Image Quality Metrics}
\label{subsubsec:imagequalitymetrics}
We define three IQMs denoted by \ent, \tv~and \cnn. 
\ent~and \tv~refer to the histogram entropy and TV of the slice images, respectively. 
We implement \ent{} following the methodology of Herbst et\,al.\,\cite{herbst2019misalignment} and \tv{} as proposed in Kingston et\,al.\,\cite{Kingston2011}.
We selected these two metrics due to their popularity in the literature. 
\ent~was found to be superior for geometry alignment in a study by Wicklein et\,al.\,\cite{wicklein2012image}.
\cnn~is our proposed method. In addition we define Ent+, Tv+ and Cnn+, each denoting an initial optimization with either Ent, Tv or Cnn followed by a fine-tuning of the annihilating trajectory with an additional optimization stage using \ent{} (for Tv+ and Cnn+) or Tv (for Ent+). 
For \ent~and \tv~the implementation of $f_\textnormal{IQM}$ is straightforward: the histogram entropy or total variation of the nine reconstructed slices are calculated. 
Following Wicklein et\,al.\,\cite{wicklein2012image} we use a bone window for the histogram calculation. Their studies showed that restricting the histogram calculation to values within a bone window improves the method's performance, because only relevant image features are captured.
The implementation of $f_\textnormal{IQM}$  for \cnn~and \cnnent~uses an additional indicator function $\vec{1}_\mathbb{M}$, where $\mathbb{M}$ describes the set of views satisfying Eq.\,\eqref{eq:softclassifier}. 
Thus, the IQM for \cnn~is defined by
\begin{equation}
f_{\textnormal{IQM}} = \regressionnet_1(\vec{y}_\textnormal{lat}(\vec{V},\vec{P}\circ \vec{C},\proj) ) \enspace \textnormal{s.\,t.} \enspace \vec{1}_\mathbb{M} |\vec{t}(\vec{m})| = \vec{0} \enspace,
\end{equation}
with $|\cdot|$ denoting element wise absolute value. 
Note that $\vec{1}_\mathbb{M}$ is not updated during the iterations.

\subsubsection{Motion Scenarios}
\begin{figure}
	\centering
\includegraphics[width=\linewidth]{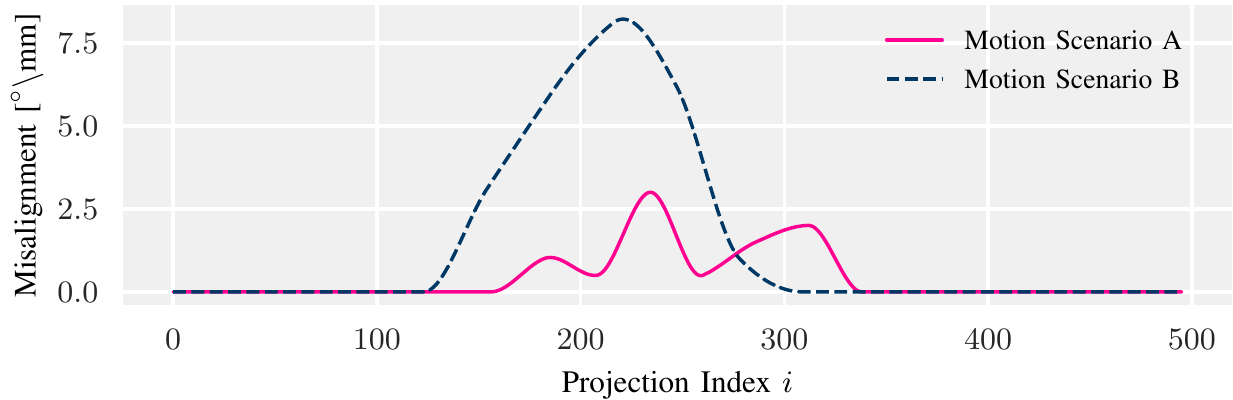}
	\caption{The motion trajectories for both motion scenarios. 
		The motion is applied respectively to the motion axis under investigation. 
		The curves are generated using 20 spline nodes and 17 spline nodes for scenario A and B, respectively.}
	\label{fig:motionshapes}
\end{figure}
We design two motion scenarios (scenario A and scenario B) differing in their motion shapes and the number of spline nodes used for both, the motion trajectory and annihilating trajectory. 
Scenario A uses 20 spline nodes for the motion curves and the same number of spline nodes for the annihilating curves. 
Scenario B uses 17 spline nodes for the motion curves and 40 spline nodes for the annihilating curves. 
The two motion amplitudes are depicted in Fig.\,\ref{fig:motionshapes}. 
For both scenarios the respective motion is applied to one of the 6 motion axes, respectively. 
In each scenario, we optimize only for the axis which is affected by the motion.
\subsubsection{Motion Estimation Results}

\def\cnn{Cnn}
\def\ent{Ent}
\def\tv{Tv}
\def\cnnent{Cnn+}
\def\noco{None}
\def\groundtruth{Gt}
\def\spacesymbol{\,}

\begin{figure*}
	\begin{minipage}[]{0.49\textwidth}
		\captionof{table}{Mean misalgnemt [\degmm] between annihilating trajectory and ground-truth trajectory for motion scenario A.}
		\label{tab:ssima}
		\resizebox{\linewidth}{!}{
			\centering
			{\def\arraystretch{1.1}\tabcolsep=-0.5pt
				\begin{tabular}{@{\extracolsep{10pt}} l c c c c c c  @{}}\hline
					& $t_x$ (ip) & $t_y$ (ip) & $t_z$ (op) & $r_x$ (op)& $r_y$ (op)&$r_z$ (ip) \\ 
					\hline
					None & 0.45 & 0.45 & 0.45 & 0.45 & 0.45 & 0.45 \\
					Ent & 0.69 & 0.20 & 0.13 & \textbf{0.10} & 0.33 & 0.33 \\
					Ent+ & 1.07 & 0.16 & 0.16 & 0.12 & 0.35 & 0.39 \\
					Tv & 0.97 & \textbf{0.12} & 0.45 & 0.32 & 0.46 & 0.69 \\
					Tv+ & 0.95 & 0.14 & 0.46 & 0.27 & 0.46 & 0.69 \\
					Cnn & \textbf{0.27} & 0.24 & 0.15 & 0.24 & 0.21 & 0.18 \\
					Cnn+ & 0.28 & 0.20 & \textbf{0.13} & 0.19 & \textbf{0.19} & \textbf{0.14} \\
					\hline 
				\end{tabular}
			}
		}
		\vspace{3mm}
		\captionof{table}{SSIM values normalized to the range [0,100] for motion scenario A. The SSIM is computed between the ground-truth reconstruction and the respective compensated reconstruction. In brackets, the SSIM is computed in a volume-of-interest. The volume of interest covers the nasal bones only.}
		\label{tab:ssima2}
		\resizebox{\linewidth}{!}{
			\centering
			{\def\arraystretch{1.1}\tabcolsep=-0.5pt
				\begin{tabular}{@{\extracolsep{10pt}} l c c c c c c  @{}}\hline
					& $t_x$  & $t_y$  & $t_z$  & $r_x$ & $r_y$ &$r_z$ \\ 
					\hline
					None & 58\spacesymbol{}(64) & 81\spacesymbol{}(89) & 75\spacesymbol{}(72) & 81\spacesymbol{}(68) & 79\spacesymbol{}(86) & 68\spacesymbol{}(66) \\
					Ent & 53\spacesymbol{}(67) & 94\spacesymbol{}(97) & \textbf{95}\spacesymbol{}(95) & 97\spacesymbol{}(97) & 89\spacesymbol{}(97) & 76\spacesymbol{}(81) \\
					Ent+ & 50\spacesymbol{}(65) & 94\spacesymbol{}(97) & 95\spacesymbol{}(\textbf{95}) & \textbf{97}\spacesymbol{}(\textbf{98}) & 89\spacesymbol{}(97) & 74\spacesymbol{}(82) \\
					Tv & 51\spacesymbol{}(66) & \textbf{97}\spacesymbol{}(\textbf{98}) & 78\spacesymbol{}(77) & 88\spacesymbol{}(85) & 82\spacesymbol{}(93) & 66\spacesymbol{}(70) \\
					Tv+ & 49\spacesymbol{}(67) & 96\spacesymbol{}(98) & 77\spacesymbol{}(78) & 90\spacesymbol{}(89) & 83\spacesymbol{}(94) & 63\spacesymbol{}(71) \\
					Cnn & \textbf{69}\spacesymbol{}(81) & 90\spacesymbol{}(95) & 92\spacesymbol{}(92) & 90\spacesymbol{}(87) & 90\spacesymbol{}(97) & 83\spacesymbol{}(87) \\
					Cnn+ & 69\spacesymbol{}(\textbf{81}) & 92\spacesymbol{}(97) & 94\spacesymbol{}(95) & 92\spacesymbol{}(92) & \textbf{92}\spacesymbol{}(\textbf{98}) & \textbf{86}\spacesymbol{}(\textbf{91}) \\
					\hline 
				\end{tabular}
			}
		}
		\begin{minipage}{\textwidth}
			\vspace{3mm}
			\includegraphics[width=\linewidth]{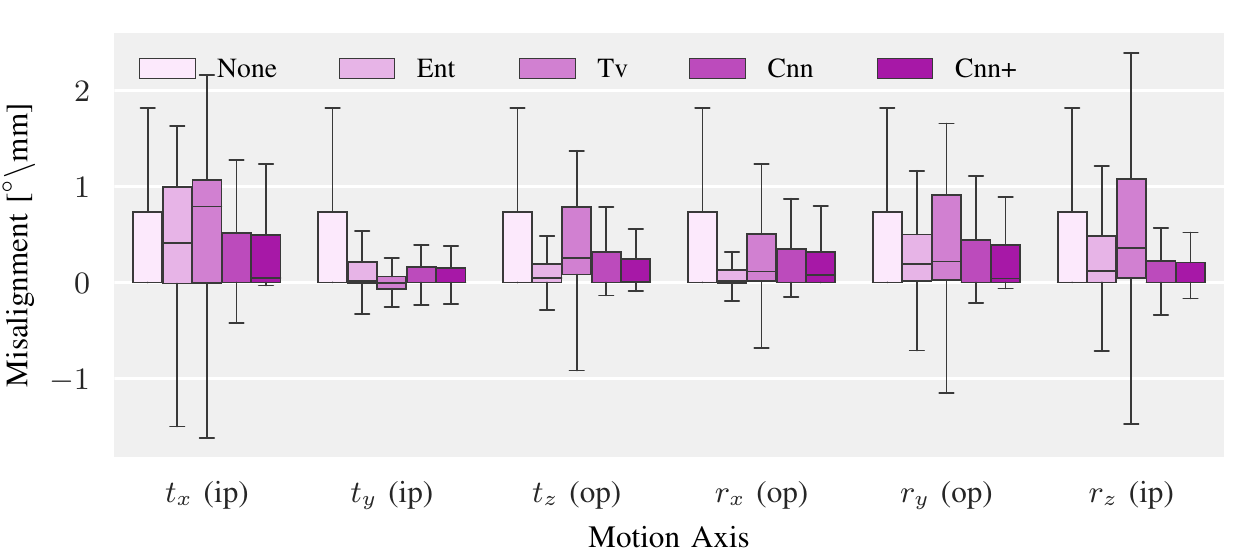}
			\captionof{figure}{Boxplot for motion scenario A, showing the misalignment of the annihilating curve to the motion curve  plotted for each of the four different IQMs used in the motion estimation benchmark.}
			\label{fig:boxplot_compensation_a}
		\end{minipage}
	\end{minipage}
	\begin{minipage}{0.2\textwidth}%
	\end{minipage}
	\begin{minipage}[]{0.49\textwidth}
		\captionof{table}{
			Mean misalgnemt [\degmm] between annihilating trajectory and ground-truth trajectory for motion scenario B.}
		\label{tab:ssimb}
		\resizebox{\linewidth}{!}{
			\centering
			{\def\arraystretch{1.1}\tabcolsep=-0.5pt
				\begin{tabular}{@{\extracolsep{10pt}} l c c c c c c  @{}}\hline
					& $t_x$ (ip) & $t_y$ (ip) & $t_z$ (op) & $r_x$ (op)& $r_y$ (op)&$r_z$ (ip) \\ 
					\hline
					None & 1.52 & 1.52 & 1.52 & 1.52 & 1.52 & 1.52 \\
					Ent & 1.70 & 1.18 & 1.24 & 1.09 & 1.32 & 1.43 \\
					Ent+ & 1.93 & 1.06 & 1.26 & 1.09 & 1.33 & 1.57 \\
					Tv & 2.01 & 1.34 & 1.54 & 1.38 & 1.55 & 1.83 \\
					Tv+ & 1.95 & 1.17 & 1.43 & 1.18 & 1.48 & 1.79 \\
					Cnn & 1.20 & 0.54 & 1.15 & 0.84 & 0.76 & 0.83 \\
					Cnn+ & \textbf{1.14} & \textbf{0.45} & \textbf{0.90} & \textbf{0.67} & \textbf{0.69} & \textbf{0.62} \\
					\hline 
				\end{tabular}
			}
		}
		\vspace{3mm}
		\captionof{table}{SSIM values normalized to the range [0,100] for motion scenario B. The SSIM is computed between the ground-truth reconstruction and the respective compensated reconstruction. In brackets, the SSIM is computed in a volume-of-interest. The volume of interest covers the nasal bones only.}
		\label{tab:ssimb2}
		\resizebox{\linewidth}{!}{
			\centering
			{\def\arraystretch{1.1}\tabcolsep=-0.5pt
				\begin{tabular}{@{\extracolsep{10pt}} l c c c c c c  @{}}\hline
					& $t_x$  & $t_y$ & $t_z$  & $r_x$ & $r_y$ &$r_z$  \\ 
					\hline
					None & 49\spacesymbol{}(49) & 65\spacesymbol{}(71) & 66\spacesymbol{}(61) & 69\spacesymbol{}(45) & 66\spacesymbol{}(68) & 54\spacesymbol{}(56) \\
					Ent & 46\spacesymbol{}(47) & 66\spacesymbol{}(76) & 66\spacesymbol{}(61) & 72\spacesymbol{}(51) & 67\spacesymbol{}(71) & 55\spacesymbol{}(55) \\
					Ent+ & 46\spacesymbol{}(47) & 68\spacesymbol{}(78) & 66\spacesymbol{}(59) & 72\spacesymbol{}(52) & 67\spacesymbol{}(71) & 55\spacesymbol{}(55) \\
					Tv & 48\spacesymbol{}(48) & 67\spacesymbol{}(74) & 65\spacesymbol{}(60) & 70\spacesymbol{}(47) & 66\spacesymbol{}(68) & 53\spacesymbol{}(55) \\
					Tv+ & 45\spacesymbol{}(46) & 67\spacesymbol{}(76) & 65\spacesymbol{}(59) & 71\spacesymbol{}(48) & 65\spacesymbol{}(69) & 52\spacesymbol{}(54) \\
					Cnn & 46\spacesymbol{}(51) & 75\spacesymbol{}(85) & 67\spacesymbol{}(66) & 73\spacesymbol{}(58) & 72\spacesymbol{}(80) & 54\spacesymbol{}(57) \\
					Cnn+ & \textbf{50}\spacesymbol{}(\textbf{54}) & \textbf{81}\spacesymbol{}(\textbf{90}) & \textbf{70}\spacesymbol{}(\textbf{67}) & \textbf{78}\spacesymbol{}(\textbf{67}) & \textbf{75}\spacesymbol{}(\textbf{84}) & \textbf{66}\spacesymbol{}(\textbf{65}) \\
					\hline 
				\end{tabular}
			}
		}
		\begin{minipage}{\textwidth}
			\vspace{3mm}
			\includegraphics[width=\linewidth]{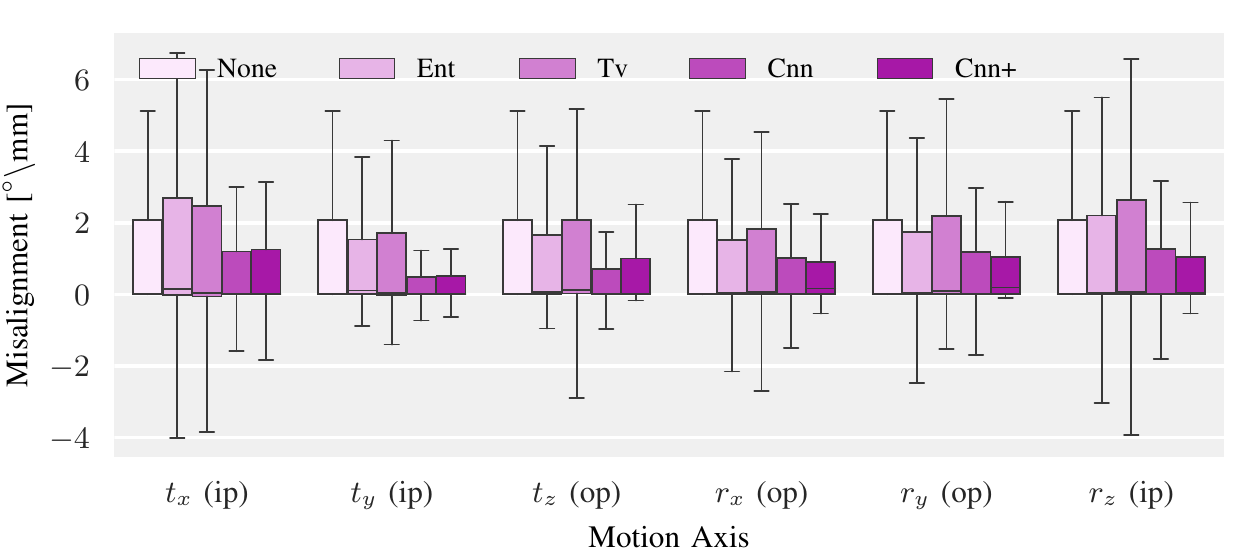}
			\captionof{figure}{Boxplot for motion scenario B, showing the misalignment of the annihilating curve to the motion curve plotted for each of the four different IQMs used in the motion estimation benchmark.}
			\label{fig:boxplot_compensation_b}
		\end{minipage}
	\end{minipage}
\end{figure*}
To quantify the performance, we measure the mismatch of the respective  motion curves and estimated annihilating curves. 
For a complete compensation, we need $\vec{M}\circ\vec{C} = \vec{1}$, which is the case if the motion curve and the annihilating curve add up to zero. 
The second metric measures the reconstruction quality by computing the SSIM of the respective motion-compensated reconstruction and the ground truth.

\begin{figure*}
	\includegraphics[width=\linewidth]{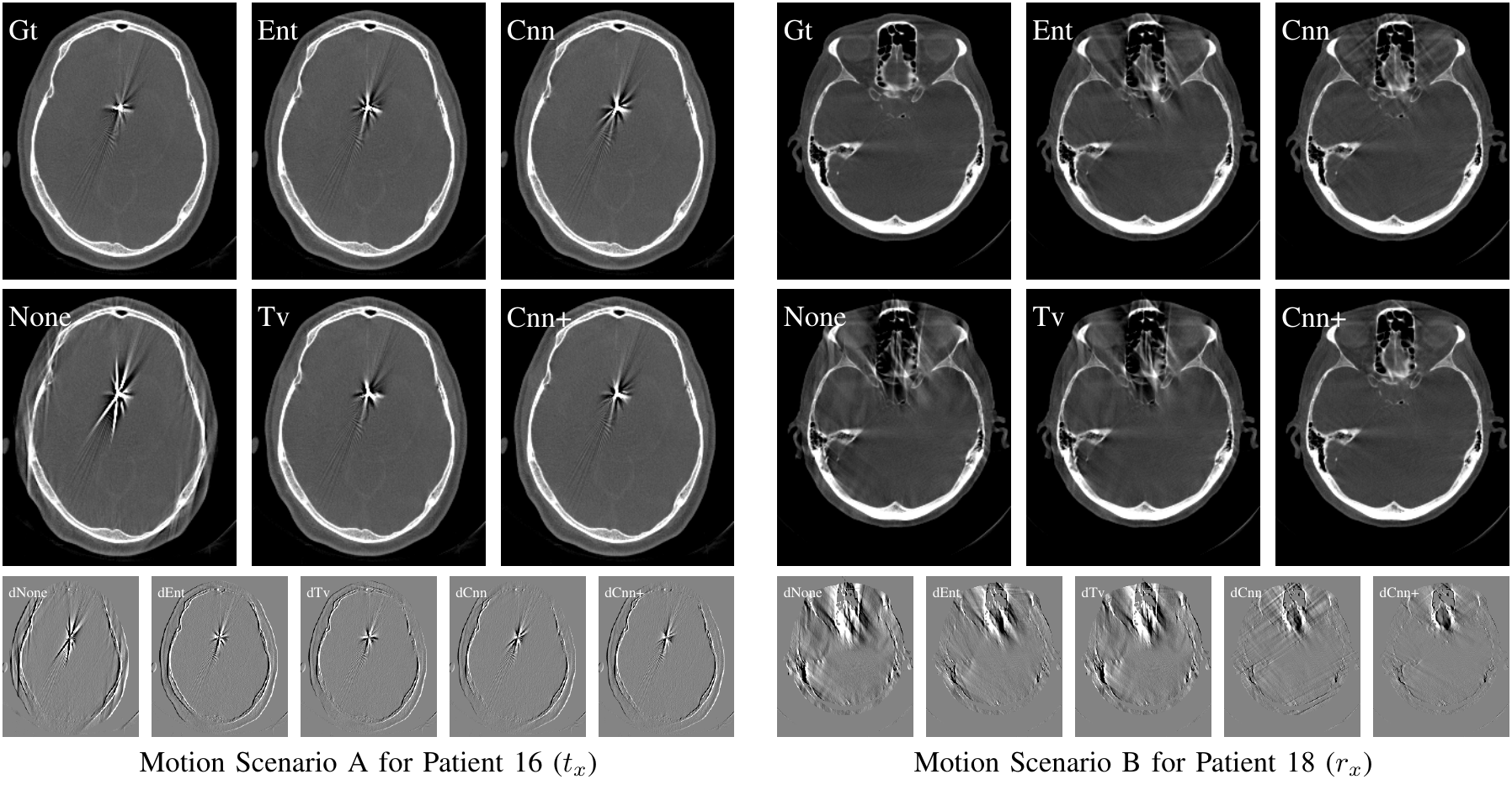}
	\caption{Selected reconstructions (HU [50-2000]) from the motion benchmark. 
		Left block: Motion scenario A using 20 spline nodes to model the annihilating trajectory, and right block: Motion scenario B using 40 spline nodes to model the annihilation trajectory. 
		The respective bottom row displays the difference images to the Ground truth (Gt). 
		The deviation of the annihilating curve to the negative motion curve is \noco~= 0.44\,mm, \ent~= 0.74\,mm, \tv~=  0.69\,mm,  \cnn~= 0.37\,mm and \cnnent~=0.45\,mm for motion scenario A, and  \noco~= 1.52\,mm,  \ent~=  0.62\,mm, \tv~=  1.23\,mm, \cnn~= 0.36\,mm and  \cnnent~= 0.31\,mm for motion scenario B. }
	\label{fig:recos}
\end{figure*}

Figure \ref{fig:boxplot_compensation_a} and Fig.\,\ref{fig:boxplot_compensation_b} show the misalignment of the motion curve for motion scenario A and B, respectively, averaged over the 4 validation and 2 test patients for Ent, Tv, Cnn and Cnn+. 
Numeric results showing the misalignment for all six metrics are displayed in Tab.\,\ref{tab:ssima} and Tab.\,\ref{tab:ssimb}  and numeric results showing the SSIM values for all six metrics are displayed in Tab.\,\ref{tab:ssima2} and Tab.\,\ref{tab:ssimb2}, respectively.
Selected reconstructions for both motion scenarios are presented in Fig.\,\ref{fig:recos}.

The proposed method performs well in both scenarios. 
In motion scenario A, the state-of-the-art methods perform similar to the network-based solution.
In the majority of cases, the network-based results are superior. However, in almost 50\% of the experiments, either \tv~or \ent~achieves the best results. A fine-tuning of the traditional metrics (Ent+, Tv+) barely improves the results and can lead to a degeneration of the performance.
The margin by which the network-based method outperforms both state-of-the-art metrics is much higher than vice-versa. 
Figure \ref{fig:recos} shows that the network-based approach is further capable of dealing with metal artifacts. 
In this case, the network without post-optimization using the entropy (\cnn) achieves the best results. 
Note, that our training set also includes patients with metal artifacts.

For scenario B, our method is constantly outperforming the state-of-the-art metrics, both in terms of SSIM and measured misalignment. 
By an additional post-optimization with the entropy-based compensation, the best results are achieved with \cnnent. 
As can be seen from Fig.\,\ref{fig:recos}, the network is capable of approximating the true motion curve but ignores small high-frequent motions. 
These motions are then eliminated by the entropy. 
However, deploying entropy alone produces mediocre results because the optimization gets stuck in local minima. 
Tv performs worst  w.\,r.\,t.\ the misalignment of the annihilating and motion curves, however, the SSIM is comparable to the entropy-based procedure.

\subsection{Motion-Affected Clinical Data}
\def\Slicea{Slice 1}
\def\Sliceb{Slice 2}
\def\slicea{slice 1}
\def\sliceb{slice 2}
\subsubsection{Data and Preprocessing}
To demonstrate the effectiveness of the proposed method in clinical practice, we apply it on a motion-affected clinical dataset. 
Similar to the acquired data used for the network training and evaluation, the patient was scanned with a clinical CBCT system (Artis Q, Siemens Healthcare GmbH, Forchheim). 
The projections were downsampled and aligned following the same procedure (i.\,e., step 1 of data generation) as presented in Sec.\,\ref{subsec:datageneration}.

\subsubsection{Motion Compensation Scheme}
We model the annihilating trajectory  with an Akima spline consisting of 20 spline nodes equally distributed over the trajectory. We adapt the optimization scheme from Sec.\,\ref{sec:optimization}.
We sequentially optimize for all six motion parameters in the following sequence ($t_z$,$t_x$,$t_y$,$r_x$,$r_y$,$r_z$). To optimize for motion we use \cnn{} and \cnnent{}.

\subsubsection{Motion Compensation Results}
\begin{figure*}
\includegraphics[width=\linewidth]{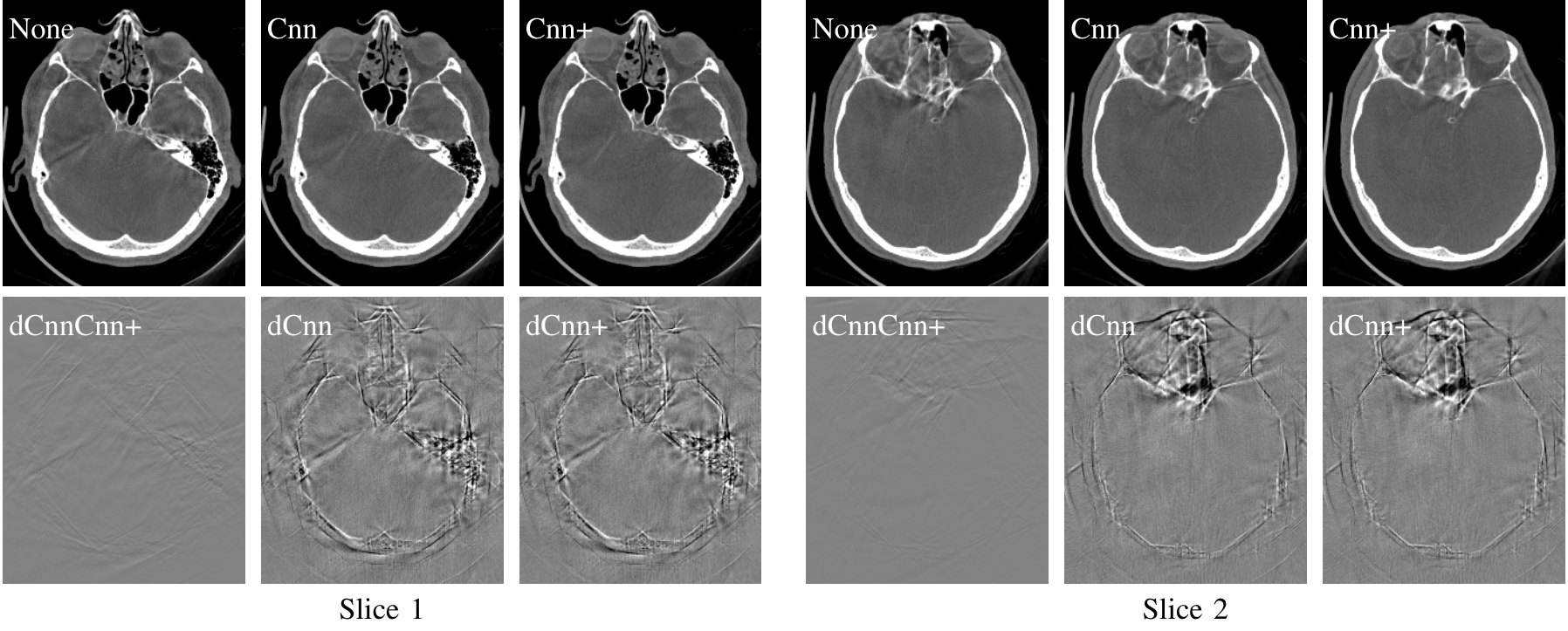}
	\caption{Top row: two reconstructed slices (HU [50-2000]) of a motion-affected clinical dataset (None) and motion-compensated data using \cnn{} or \cnnent, respectively. Bottom row: difference images of motion compensated and motion-affected reconstruction (d\cnn, d\cnnent) as well as difference image of motion compensated reconstructions using \cnn{} and \cnnent{}, respectively (d\cnn{}\cnnent{}). Display windows within each row are equal.}
	\label{fig:clinicalresults}
\end{figure*}

Figure \ref{fig:clinicalresults} displays reconstructed slice images from the motion-affected clinical dataset (\noco) as well as motion-compensated reconstructions ({}\cnn, \cnnent). 
As ground-truth reconstructions are not available, only a qualitative inspection is possible. 
In slice 1 we observe motion artifacts especially at the boarders of the temporal bones as well as near the nasal cavities and ethmoid bone. 
The anatomy contours can be well recovered using \cnn{} or \cnnent. As can be observed from the difference images (d\cnn, d\cnnent), streaks at the bone contours are eliminated. In slice 2 the motion artifacts are severe in the orbital bone structures. The streak artifacts are reduced in both motion-compensated reconstructions, restoring the homogeneous regions. A small residual motion is still observable with the motion-compensated reconstructions, however, the image quality could be improved substantially.
From the difference images between the \cnn-based and \cnnent{}-based compensated reconstructions (d\cnn\cnnent) we see that the entropy-based fine-tuning barely affects the reconstruction quality.

\section{Discussion}
\label{sec:discussionandconclusion}
We propose an appearance learning approach that can be deployed for image-based motion compensation. 
For that purpose, we devise a framework that learns the mRPE as well as vRPEs from reconstructed slice images. 
Exact computation of the RPE allows for geometric calibration for high-quality CBCT \cite{strobel2003improving}. Hence, given a 100\% network accuracy, minimizing the network-predicted mRPE would yield highly accurate motion parameters.
The axis with lowest accuracy in predicting the mRPE is also the axis with lowest performance in the motion compensation benchmark.

We further show that we can learn general features applicable to all three types of translations and rotations. 
The learned features are even less dependent on the motion axis than traditional methods. 
For example, Tv shows superior results in compensating translation along the y-axis as observable from motion experiment A. 

Autofocus methods are characterized by optimizing an IQM in the reconstruction domain. 
Inherently, information from which part of the trajectory a misalignment is expected can only be deduced from the gradients within the optimization. 
We aim to overcome this by learning an initial estimate about the distribution of the expected motion. 
The view-wise prediction can be used as a soft-classifier to steer the optimization. The FN range is close to $0$\%, ensuring that the optimization cannot be worsen by applying the soft-classifier for optimization steering.

We use randomly generated motions that are limited in their frequency due to the deployed splines. However, the network is capable to generalize to unseen motion frequencies. In motion scenario B we compensate the motion with a spline controlled by 40 spline nodes, whereas in training, only motion patterns generated with 20 spline nodes were shown to the network. Hence, the network is capable to generalize to higher frequency motion patterns. However, we can observe, that very small but high-frequent motion artifacts are barely accounted for by the network (see Fig.\,\ref{fig:recos}, scenario B). These types of motion patterns where not part of the network training. Besides those small motion artifacts, the overall motion trajectory can be estimated well by the network. 
This property is synergistic with traditional IQMs. After fine-tuning with the entropy-based IQM the fine motion artifacts are eliminated.

Traditional IQMs measure the artifact strength by a limited set of image features. If the reconstruction is not corrupted by motion, the image shows homogeneous soft-tissue areas and clear bone-boundaries. This results in a low TV value and low histogram entropy. Motion corrupts the homogeneous regions and blur bone-edges increasing the histogram entropy and TV value. 
However, the image features recognized by TV and entropy are not directly linked to the patient motion strength and therefore are susceptible to local minima. Thus, both metrics are successful if the motion is small but fail if the motion is large.	
This is shown by the experiments, where both metrics, \tv{} and \ent{}, perform well in scenario A. For larger motions as apparent in motion scenario B, both metrics fail. In those scenarios, the learned metric \cnn{} outperforms the traditional methods in all six experiments.

Besides being only trained on synthetically generated motion, the network generalizes well to real clinical motion. We demonstrate this using a clinical motion-affected scan. Due to the rigid structure of the motion, a transformation of the object can be equivalently described by a static object and a transformation of the system geometry. This allows realistic generation of motion artifacts from artifact-free CBCT acquisitions.

\section{Conclusion and Outlook}
Our proposed method can be used in a variational manner for image-based autofocus techniques. 
The result is always based on the acquired raw-data and ensures data integrity. 
This is a strong advantage to all other learning-based approaches found in our literature review. 
Current learning-based approaches perform an image-to-image translation, without any guarantee for the consistency with the acquired raw data. 
In contrast, using the proposed method the images are always reconstructed from the raw data minimizing the risk for generating clinical images leading to improper diagnosis.

The experiments show that motion artifacts can be learned by a neural network and  that our learning-based approach can outperform state-of-the-art IQMs in a motion estimation benchmark. 
We  devised the approach based on the FDK algorithm and artificial motion. 
Using a motion-affected clinical dataset, we further demonstrate that the method translates to real clinical motion.
The FDK is suitable for autofocus approaches \cite{sisniega2017motion,Kingston2011} due to its computational efficiency. 
A possible extension, however, would be a reconstruction algorithm, capable of reconstructing arbitrary trajectories \cite{Defrise1994}. 
The FDK assumes two fundamental properties: (1) homogeneous object in the direction perpendicular to the acquisition plane and (2) equally sampled trajectories along an arc. 
If any of those assumptions are not met, the reconstruction reveals cone-beam artifacts or intensity inhomogeneities. 
Therefore, it can only compute approximate solutions for motion compensation. 

Although our experiments are tailored for head CBCT, the concept is neither limited to rigid head motion nor to transmission imaging. 
By replacing the filtered back-projection with the inverse model for MRI~---~e.\,g., non-uniform Fourier transform \cite{fessler2007nufft}~---~the approach can be directly trained for propeller  trajectories in MRI. 
By additionally replacing the RPE-based regression metric with an appropriate metric (e.\,g., energy of a spline deformation field), also Cartesian sampled MRI can be tackled. 
Similar strategies are thinkable for PET.
\\
\\{}
\textbf{Disclaimer:} The concepts and information presented in this article are based on
research and are not commercially available.

\bibliographystyle{IEEEtran}

\bibliography{99lib}

\end{document}